\documentclass[aps,prl,secnumarabic,
twocolumn,groupedaddress,showpacs,amsmath,amssymb]{revtex4}
\usepackage[T2A]{fontenc}
\usepackage[cp1251]{inputenc}
\usepackage[english,russian]{babel}

\usepackage{bm}

\begin{document}
\title{Instabilities of one- and two-dimensional degenerate atomic Fermi gas against a
long-wave perturbation in optical lattice.}
\author{L.~A.~Manakova}
\email[]{manakova@kurm.polyn.kiae.su} \affiliation{ RRC Kurchatov Institute, Kurchatov
Sq.\ 1, 123182 Moscow, Russia}

\vspace{0.5cm}

\vspace{0.5cm}

\begin{abstract}
A mechanism of both formation of peaks in the density of states near the Fermi surface
and phase instabilities of nearly ideal degenerate Fermi gas in low-dimensional
optical lattices is proposed. According to this mechanism, peak formation is caused by
the quasi-classical quantization of the one- and two-dimensional fermionic spectrum in
the neighborhood of its extremal points under interaction with an long-wave periodical
perturbation. The new spectra result in  the instabilities with respect to spontaneous
formation of an equilibrium superstructure. In the one-dimensional case this happens
for low enough numbers of fermionic atoms. As a result of such transition, fermions
become localized (a transition of the metal-insulator type). In the two-dimensional
system the transition is possible for a nearly half-filled band. In this case fermions
are localized in the wave direction only. It is briefly discussed the possible
influence of the results obtained in the paper on the superfluid transition
temperature in high anisotropic lattices possessing quasi-(one,two)-dimensional
subsystems of fermionic atoms.

\end{abstract}
\maketitle

\section{Introduction}
A quantum degenerate dilute gas of fermionic $^{40}K$ atoms has been trapped and
cooled down to a fraction of the Fermi temperature $T_F$ in the works \cite{MJ},
\cite{MJ1}. Magneto-optical trapping of $^{6}Li$ atoms has been realized in the work
\cite{MFS}. These results enable to explore quantum degenerate Fermi systems with
controlled particle interactions. In the atomic traps one can study the Fermi systems
with almost any number of particles and interaction strength. The latter is achieved
by tuning Feschbach resonance \cite{RCC}. Predicted phenomena for an interacting
atomic Fermi gas include zero sound and hydrodynamic excitations \cite{YH}, \cite{BC},
suppression of elastic and inelastic collisions \cite{MJ}, \cite{MJ1}, \cite{F}, and
the prospect of a state with pairwise correlations, analogous to Cooper pairing of
electrons (see, for instance, \cite{SHS}, \cite{HKC}, \cite{HM1}, \cite{BM}).
Theoretical studies of the superfluidlike (BCS) state have concluded that this state
will occur at very low temperature and will be difficult to observed experimentally.
In the works \cite{TZ}, \cite{PRT}, \cite{PRT1}, \cite{R} there was proposed a method
to detect the existence of the BCS state by using interactions of atoms with
off-resonant laser light.

However, a trapped gas of fermionic atoms that is quantum degenerate but uncondensed
is interesting in its own right. Quantum statistic radically changes the collision
properties, spatial profile, and off-resonant light-scattering properties \cite{MJ},
\cite{BC} \cite{MPJ}, \cite{MKH}. In the initial works \cite{MJ1}, \cite{MFS} a nearly
ideal Fermi gas has been produced. Since Fermi statistics prohibits s-wave collisions
between identical particles, for fermionic atoms the dominant contribution must be p
wave. However, the p-wave collision cross section is suppressed at low temperature due
to the centrifugal barrier. The threshold energy is equal to $E_{th}(l=1)\gg T_F$
\cite{MJ}. Thus, below this energy the elastic collision rate in a gas of identical
fermionic atoms will rapidly vanish. A quantum-statistical suppression of s-wave
interactions between identical fermions makes  an ultracold gas of fermionic atoms an
excellent realization of an ideal gas.

If one produces an atomic Fermi gas in two spin states (that are, for example,
magnetic sublevels $m_f=9/2$ and $m_f=7/2$ in the $^{40}K$ $f=9/2$ ground state),
where both species are quantum degenerate, then s-wave interactions are exhibited. The
emergence of quantum degeneracy is observed through measurements of the equilibrium
thermodynamic properties of the two-component gas \cite{MJ1}. In the work \cite{MPJ}
there are presented measurements of the equilibrium thermodynamic properties and
collision dynamics of an interacting degenerate Fermi gas in two spin states. As is
shown, Pauli blocking of collisions occurs due to the quantum statistic. For instance,
at quantum degenerate temperatures allowed spin-exchange collisions can be suppressed
through final-state occupation. The specific dynamics of both elastic and inelastic
collisions has a decisive role for properties of an degenerate atomic Fermi gas.

It is worth noting that a quantum degenerate gas of fermionic atoms in optical
lattices is relatively unexplored theoretically as well as experimentally. While this
situation can provide a richer spectrum of phenomena arising in both interacting and
even ideal Fermi system. In particlar, there appear possibilities of changing the
dominant properties of an atomic Fermi gas that are specified for lattices only.

{\bf 2.} The present paper focuses on a mechanism of formation of peaks in the density
of states near the Fermi surface and phase instabilities of nearly ideal Fermi gas in
low-dimensional optical lattices. The peaks and instabilities are generated by a
long-wave perturbation in an optical lattice.

According to the mechanism proposed in what follows, peak formation is caused by the
quasi-classical quantization of the one- and two-dimensional fermion spectrum in the
neighborhood of its extremal points under interaction with an long-wave perturbation.
Such perturbation can be generated either by polarization (as well as magnitization)
wave or by the long-wave deformation of the optical lattice. These perturbations are
produced either by the external drive or by the long-wave fluctuations of an order
parameter above the transition point.  We show that a highly nonequidistant discrete
spectrum forms in the neighborhood of the extremal points provided the wave vector $k$
of a periodical perturbation satisfies the condition
\begin{equation}
k\ll (g/t)^{1/2}\ll 1 \label{eq:f1}
\end{equation}
where $g$ and $t$ are the perturbation magnitude and bandwidth. Near a saddle point of
the spectrum of a square lattice, the spectrum resembles a collection of Landau levels
in a magnetic field,but in contrast to the latter is nonequidistant. In the
one-dimensional case the spectrum has a logarithmic peak in the density of states at
the boundary separating the discrete spectrum from the continuous near the bottom or
top of the band. In the two-dimensional case the spectrum generated by the long-wave
perturbation provides a flat maximum in the density of states near a saddle point. The
peak widths $\sim g$ is much smaller than the bandwidth of the atomic gas.

In the both cases the spectrum results in  the instabilities with respect to
spontaneous formation of an equilibrium superstructure ( or, what is the same, an
equilibrium wave of polarization, magnitization or deformation). In one-dimensional
case this happens for a fairly low fermion number, when $\mu<\mu_c\ll t$. In the
two-dimensional case the transition is possible only provided the Fermi energy is
exceptionally close to a saddle point. As a result of such transition in the
one-dimensional case, the fermions become localized (a transition of the
metal-insulator type). In the two-dimensional case the fermions are localized in a
single dimension, that should lead to a sharp anisotropy of transport properties of
the system.

In conclusion, it is discussed the possible influence of the results obtained in the
paper on the superfluid transition temperature in high anisotropic lattices possessing
quasi-(one,two)-dimensional subsystems of Fermi atoms.

\section{Quasiclassical spectra and the densities of states}

{\bf 1.} We start with the two-dimensional case. For a static wave pointing in an
arbitrary chosen direction, $g(x,y)=g\cos(k_1 x+k_2 y)$, the Hamiltonian of the system
has the form

\begin{equation}
 H=\varepsilon_0(p_x, p_y)-g\cos(k_1 x+k_2 y).
\label{eq:f3}
\end{equation}

The shape of the spectrum change drastically in the neighborhood of the values
$p_{xm}$ and $p_{ym}$ that satisfy the condition

\begin{equation}
 k_1\frac{\partial\varepsilon_0}{\partial
 p_x}+k_2\frac{\partial\varepsilon_0}{\partial p_y}=0,
\label{eq:f2}
\end{equation}
This condition specifies the region of anomalous condensation of energy values. Hence,
the $g(x,y)$ perturbation leads near the values defined by Eq.(\ref{eq:f3}) to
transformation7 of a large number of levels.

For the sake of simplicity7, let us consider a square lattice with
$\varepsilon_0(p_x,p_y)=-2t(\cos p_x+\cos p_y)$. In this case the condition
(\ref{eq:f2}) determines, for instance, the saddle points $(0,\pi)$, $(\pi,0)$ of the
spectrum. In the neighborhoodof saddle points the spectrum corresponds to a nearly
half-filled band.

Near the saddle point $(0,\pi)$ the Hamiltonian (\ref{eq:f3}) is diagonalized in the
new canonical variables:
\begin{equation}
\begin{split}
&q_1=\sqrt{2}k[p_x\sin\varphi-p_y\cos\varphi];\\
&q_2=\sqrt{2}k[p_x\sin\varphi-p_x\cos\varphi];\\ &X=k(x\cos\varphi+y\sin\varphi);\\
&Y=k(x\cos\varphi-y\sin\varphi),
\end{split}
\label{eq:f4}
\end{equation}
where we introduce the angle $\varphi$ that specifies the direction of the wave:
$k_1=k\cos\varphi$, $k_2=k\sin\varphi$. A wave along one of the principal axes
corresponds to $k_2=0$ and $\varphi=0$. As a result of the transformation specified by
Eqs. (\ref{eq:f4}), the Hamiltonian  (\ref{eq:f3}) near the saddle point becomes
\begin{equation}
H=-4tk^{2}(q_1^2-q_2^2)\cos2\varphi-2g\cos X; \label{eq:f5}
\end{equation}
here $\cos2\varphi<0$. The Schr$\ddot{o}$dinger equation takes the form

\begin{equation}
\Big[(-\tilde{t}q_1^2+2g\cos X) + \tilde{t}q_2^2\Big]\Psi=\varepsilon\Psi,
\label{eq:f6}
\end{equation}
where $\tilde{t}=tk^2\cos2\varphi$. The eigenvalues of Eq.(\ref{eq:f6}) have the form
\begin{equation}
\varepsilon_{(0,\pi)}(n,q_2)=\varepsilon_n+\tilde{t}q_2^2, \label{eq:f7}
\end{equation}
where $\varepsilon_n$ is the spectrum in the periodical potential in the $X$
direction.

In the neighborhood of the second saddle point (or, what is the same, at
$\cos2\varphi>0$) the spectrum is inverted,
\begin{equation}\label{eq:f8}
\varepsilon_{(\pi,0)}(n,q_2)=-\varepsilon_{(0,\pi)}(n,q_2).
\end{equation}

In the $X$ direction (or, equivalently, for each fixed value of $q_2$) the fermionic
spectrum is determined by Mathieu's equation that is found from (\ref{eq:f6}) by means
of the usual quantization rule, $q_1\rightarrow -i\partial/\partial X$
\begin{equation}\label{eq:f9}
\left(\tilde{t}\partial^2/\partial X^2+2g\cos X\right)\Psi_{q_2}=\varepsilon\Psi_{q_2}
\end{equation}
For $\varepsilon>2g$ this equation has a continuous spectrum.In the region $-2g <
\varepsilon < 2g$ Eq.(\ref{eq:f9}) has narrow allowed bands. If we irnore the
bandwidths, we can speak of a discrete spectrum in wells whose number is $L/2\pi$
where $L$ is the size of the system in the $X$ direction.

Provided the condition $$n_c=\frac{8}{\pi}\left(\frac{g}{|\tilde{t}|}\right)^{1/2}\gg
1$$ is met, the number of levels in a well is large and the discrete spectrum is
specified by the Bohr-Sommerfeld formula:
\begin{equation}
\begin{split}
 &\quad n(\varepsilon_{n})=
 =\oint\frac{d X}{2\pi}\Big[\frac{1}{|\tilde{t}|}
 \left(\varepsilon_{n}+2g\cos X\right)\Big]^{1/2} \\
 &=n_c\Big[E(\kappa)-(1-\kappa^2)K(\kappa)\Big];
\end{split}
\label{eq:f10}
\end{equation}
where $\kappa^2=(\varepsilon_{n}+2g)/4g$, $K(\kappa)$ and $E(\kappa)$ are complete
elliptic integrals of the first and second kinds. It is worth noting that the quantity
$n(\varepsilon_{n})$ changes within the region of $n(\varepsilon_{n})\leq n_c$

The level separations in (\ref{eq:f10}) are given by the following expression
\begin{equation}
\omega(\varepsilon_{n})=\frac{d\varepsilon_n}{d n}=\frac{\pi\omega_m}{2K(\kappa)},\;\;
\omega_m=2(|\tilde{t}|g)^{1/2}. \label{eq:f11}
\end{equation}
At $(g/|\tilde{t}|)^{1/2}\gg 1$ we have the relation $\omega_m\ll 2g$. The quantity
$\omega_m$ determines the maximum splitting of the levels in the field of the static
wave.

The spectrum (\ref{eq:f7}) with $\varepsilon_n$ taken from (\ref{eq:f10}) resembles
the spectrum of an electron in a magnetic field, but branches of the parabola go
downward from $q_2$. The separation of the $\varepsilon_n$-levels decreases rapidly as
$n$ increases, and above the line $\varepsilon=2g+\tilde{t}q_2^2$ the spectrum becomes
that of free two-dimensional motion.

Combining the condition $q_{1max}=\sqrt{g/tk^2|\cos2\varphi|}\ll
p_{1m}=\pi/2k\cos(\varphi-\pi/4)$ with the requirement that $\omega_m\ll g$, we arrive
at the condition for the applicability of solutions obtained from the model
(\ref{eq:f5}):
\begin{equation}
k^2|\cos2\varphi|\ll \frac{g}{t}\ll \Big|\tan(\varphi-\pi/4)\Big|. \label{eq:f12}
\end{equation}
At $\varphi=0$ this condition coincides with (\ref{eq:f1}), while it becomes much more
stringent as $\varphi\rightarrow \pi/4$. As condition (\ref{eq:f12}) implies, as
$\varphi$ grows the region of quasilocalized states becomes narrower. We see that the
region with the discrete spectrum is the biggest when the wave is directed along one
of the principal axes ($\varphi=0$ in the aboveobtained expressions).

{\bf 2.} In the one-dimensional case with a static wave along the direction of the
chain, the Schr$\ddot{o}$dinger equation near the extremal points of the spectrum
takes the form
\begin{equation}
\Big(-t\frac{d^2}{dx^2}+2g\cos kx\Big)\Psi=\varepsilon\Psi, \label{eq:f13}
\end{equation}
where $x$ is the coordinate along the chain. The discrete spectrum exists within the
region of $-2g<\varepsilon<2g$ and is specified by the Eq.(\ref{eq:f10}) with
\begin{equation}
n_c=\frac{k_c}{k},\;\;\;k_c=\frac{8}{\pi}\left(\frac{g}{t}\right)^{1/2},\;\;\;
|\tilde{t}|=k^2 t, \label{eq:f010}
\end{equation}
where $k\ll k_c$.
 Respectively, the level separation is given by the
expression (\ref{eq:f11}) with $\varphi=0$ and $\tilde{t}$ defined in
Eq.(\ref{eq:f010}). In the continuous spectrum region ($\varepsilon>2g$)
\begin{equation}
n(\varepsilon)=\frac{4L}{\pi^2} \left(\frac{g}{t}\right)^{1/2}\kappa E(\kappa^{-1}).
\label{eq:f14}
\end{equation}
Unlike the two-dimensional case these results refer directly to a low occupancy of the
initial band $\varepsilon_0(p)=-2t\cos p$. In the case of an almost completely filled
initial band, the quantization of the states in a periodical potential can be obtained
via transition to the hole representation.

{\bf 3.} In the one-dimensional case we arrive at the expression for the density of
states in the discrete spectrum region combining Eqs.(\ref{eq:f10}) and
(\ref{eq:f010}):
\begin{equation}
\rho_d(\varepsilon)=\frac{kL}{2\pi}\frac{dn}{d\varepsilon}=
\frac{L}{\pi^2}\frac{K(\kappa)}{(gt)^{1/2}}. \label{eq:f15}
\end{equation}
Using Eq.(\ref{eq:f14}), we obtain the density of states in the continuous spectrum
region
\begin{equation}
\rho_c\varepsilon)=\frac{L}{2\pi^2} \frac{\kappa^{-1} K(\kappa^{-1})}{(gt)^{1/2}}.
\label{eq:f16}
\end{equation}
As Eqs.(\ref{eq:f15}) and (\ref{eq:f16}) imply, there emerges a new logarithmic
singularity at the boundary separating the discrete and continuous spectra:
\begin{equation}
\rho_{d,c}(\varepsilon)\sim \omega_m^{-1}\ln|1-\varepsilon/2g|^{-1};\;\;\;
\varepsilon\rightarrow 2g^{\pm}. \label{eq:f17}
\end{equation}
Thus, near a bottom (top) of the initial one-dimensional band the root singularity in
the density of states turns into the logarithmic one at $\varepsilon\rightarrow 2g$.
The width of the logarithmic peak is of the order of $g\ll t$.

For the two-dimensional case the DOS has a logarithmic singularity on the line $\vert
\epsilon \vert=0$ in the absence of a long-wave perturbation in Eq.(\ref{eq:f3}):
\begin{equation}
\rho_{0}(\epsilon )=\rho_{00}\ln(\frac{t}{\vert \epsilon  \vert}) \;
\;\rho_{00}=\frac{L^{2}}{(2\pi )^{2}t}. \label{eq:f017}
\end{equation}
The perturbation smooths out this singularity. Indeed, if the wave is sufficiently
long [Eq. (1)] electron motion is quasiclassical both in the continuous spectrum
region and in the region of states localized along the wave direction. Hence, for an
arbitrary directed wave the DOS is given by the integral:
\begin{equation}
\begin{split}
&\rho(\epsilon )\\&=\int \frac{dxdp_{x}}{2\pi }\int \frac{dydp_{y}}{2\pi }
\delta[\epsilon  + 2t(\cos p_{x} + \cos p_{y})- 2g\cos ({\bf k}{\bf r})]
\end{split}
\label{eq:f018}
\end{equation}
Integration over momenta and along the direction perpendicular to the wave direction
yields
\begin{equation}
\begin{split}
\rho(\epsilon )&=(1/\pi )\int_{0}^{\pi }dX \rho_{0} (\epsilon + 2g\cos X)\\
&=\rho_{00}\ln(t/g),\;\;\;  \vert \epsilon \vert <2g,\\
 & =\rho_{00}\ln\Big[\frac{t}{\vert \epsilon  \vert } + \frac{g^{2}}{\epsilon
^{2}}\Big], \;\;\; \epsilon\gg 2g.
\end{split}
\label{eq:f18}
\end{equation}
We see that the DOS in the layer between $-2g$ and $2g$"freezes" at the constant level
approximately $\ln(t/g)$ high. For $\vert \epsilon  \vert=2g$ the DOS has a break in
the slope.

Using Eqs.(\ref{eq:f10}),(\ref{eq:f11}), one can easily understand why the density of
states in (\ref{eq:f018}), (\ref{eq:f18}) is direction-independent. The reason is that
as the direction changes, an increase in the number of levels $n_c$ in a well
corresponds to a decrease in the number of wells that "fit" in to length $L$. As a
result, the dependence on direction in the density of states vanishes. Clearly, the
density of states begins to depend on ${\bf k}$ when $k^2\approx g/t$ and the
quasiclassical conditions (\ref{eq:f1}) and $\omega_m\ll g$ are break down.

\section{Instabilities to formation of a equilibrium superstructure}

Knowing the density of states, we can easy find the thermodynamic potential of the
system and study the stability of the system at $T=0$. In the one-dimensional case the
problem can be solved exactly, while in the two-dimensional case only with logarithmic
accuracy.

{\bf 1.} We start with the one-dimensional case. If the chemical potential lies below
the upper edge of the discrete spectrum (\ref{eq:f10}), $-2g<\mu<2g$, using
Eq.(\ref{eq:f15}), we arrive at the following expressions for the particle number
$N(\mu)$, the thermodynamical potential $\Omega_g(\mu)$, and the energy
$\mathcal{E}=\Omega+\mu N$ (here $\mu$ is measured from the bottom of the initial
band, and the length $L$ is assumed equal to unity):
\begin{equation}
\begin{split}
&N(\mu)=\int\limits^{\mu}_{-2g} d\varepsilon \rho_d(\varepsilon)=\frac{k}{2\pi}n(\mu)=
N_c\left[E(\kappa)-(1-\kappa^2)K(\kappa)\right];\\
&\Omega_g(\mu)=-4\mathcal{E}_c\left[(4\kappa^2-2)E(\kappa)+
(3\kappa^2-2)(\kappa^2-1)K(\kappa)\right];\\
&\mathcal{E}_g=\mathcal{E}_c\left[(2\kappa^2-1)E(\kappa)+
(6\kappa^2-1)(\kappa^2-1)K(\kappa)\right];\\
&N_c=\frac{4}{\pi^2}\left(\frac{g}{t}\right)^{1/2},\;\;
\mathcal{E}_c=\frac{2g}{9}\cdot N_c; \;\;\;\kappa^2=\frac{\mu+2g}{4g}.
 \label{eq:f19}
\end{split}
\end{equation}

If the chemical potential lies in the continuous spectrum region ($\mu>2g$), we have
\begin{equation}
\begin{split}
&N(\mu)=\int\limits^{\mu}_{2g} d\varepsilon \rho_c(\varepsilon)=N_c\kappa
E(\kappa^{-1});\\
&\Omega_g(\mu)=-4\mathcal{E}_c\kappa^3\left[(4-2\kappa^{-2})E(\kappa^{-1})+
(\kappa^{-2}-1)K(\kappa^{-1})\right];\\
&\mathcal{E}_g=\mathcal{E}_c\kappa^3\left[(2-\kappa^{-2})E(\kappa^{-1})+
4(1-\kappa^{-2})K(\kappa^{-1})\right]. \label{eq:f20}
\end{split}
\end{equation}
In both cases,
\begin{equation}
\left(\frac{\partial\Omega_g}{\partial g}\right)_{\mu}=\frac{(3/2)\Omega+ \mu N}{g}.
\label{eq:f21}
\end{equation}
For the sake of comparison, we give the expressions in the case of the unperturbed
spectrum:

\begin{equation}
\begin{split}
&\rho_{0}(\mu)\sim \frac{1}{(\mu t)^{1/2}},\;\;\;N_{0}(\mu)\sim
\left(\frac{\mu}{t}\right)^{1/2},\\ &\Omega_0=-\frac{2}{3}\mu
N_{0}\;\;\;E_0=\frac{1}{3}\mu N_{0}. \label{eq:f22}
\end{split}
\end{equation}

It is worth noting that the chemical potential $\mu$ is determined by the equation
$$\int\limits^{\mu}_{-2g} d\varepsilon \rho_d(\varepsilon)= \int\limits^{\mu_0}_{0}
d\varepsilon \rho_0(\varepsilon)$$ or, what is the same,
$$\frac{\pi}{4}\left(\frac{\mu_0}{g}\right)^{1/2}=
\left[E(\kappa)-(1-\kappa^2)K(\kappa)\right]\equiv f_{\mu}(\kappa).$$ At $-2g\leq
\mu\leq 2g$ the function $f_{\mu}(\kappa)$ monotonically increases from 0 to 1. As a
result, the equation for $\mu$ has solutions under condition $(\mu_0/g)^{1/2}\leq
(4/\pi)$.

The variations of all thermodynamic quantities due to the wave are the greatest when
$\mu\leq 2g$ and, with a further increase in the chemical potential, decrease in the
following manner:
\begin{equation}
\begin{split}
&\delta N(\mu)= N(\mu)-N_0(\mu)\sim \mu^{-3/2};\\
&\delta\Omega=\Omega_g(\mu)-\Omega_0\sim \mu^{-1/2};\;\;\delta\Omega\sim \delta
\mathcal{E}. \label{eq:f23}
\end{split}
\end{equation}

It follows from Eqs.(\ref{eq:f19})-(\ref{eq:f22}) that $\Omega_g<0$ over the entire
range of $g$ from $0$ to $\infty$. What is very important is that for all $g$ values
the thermodynamic potential $\Omega_g$ is decreasing function of $g$; i.g.,
\begin{equation}
\left(\frac{\partial\Omega_g}{\partial g}\right)_{\mu}<0. \label{eq:f24}
\end{equation}
Thus, {\bf the system is thermodynamically unstable toward the growth of the amplitude
of the periodical perturbation}.

At $g\gg \mu$ the thermodynamic potential $\Omega_g(\mu)$ in Eq.(\ref{eq:f19})
decreases as $-g^{3/2}$ with increasing the amplitude $g$. When the latter becomes
large enough, the terms proportional to $g^2$ must take into account in the
thermodynamic potential. These terms are generated, for instance, by an interaction
between polarizations (or between spins) of different atoms.

{\bf 2.} Thus, to obtain equilibrium values of $g$ one can write the phenomenological
Landau functional in the form
\begin{equation}
\mathcal{F}=\Omega_g(\mu,g)+\alpha g^2, \label{eq:f25}
\end{equation}
where the term $\alpha g^2$ describes the "rigity" of the system against the increase
in $g$.

The equilibrium $\bar{g}$ values are found from the condition for the minimum of the
functional (\ref{eq:f25}), i.g., they are the solutions of the equation
\begin{equation}
\left|\frac{\partial\Omega_g}{\partial g}\right|_{g=\bar{g}}= \alpha\bar{g}.
\label{eq:f26}
\end{equation}
Eqs. (22)-(24) imply that
\begin{equation}
\begin{split}
&\left|\partial\Omega_g/\partial g\right|\sim g/(\mu t)^{1/2}\;\; \text{at}\;\; \mu\gg
g;\\ & \left|\partial\Omega_g/\partial g\right|\sim (g/t)^{1/2}\;\;\; \text{at}\;\;
\mu\ll g;
\end{split}
\label{eq:f27}
\end{equation}
$$\left|\partial^2\Omega_g/\partial g^2\right|\sim \ln|2g-\mu|^{-1};\;\;\; at
\;\;\;|\mu-2g|\rightarrow 0.$$ The domains of $g$ values to the right and left of the
$2g=\mu$ point correspond of the discrete and continuous spectra, respectively. The
tangent to the $\left|\partial\Omega_g/\partial g\right|$ curve at the zero point is
\begin{equation}
\alpha_0=\frac{1}{\pi}(\mu t)^{-1/2}. \label{eq:f28}
\end{equation}
Solution of Eq.(\ref{eq:f26}) yields the following results.

At $\mu>0$ and $\alpha<\alpha_0$ the state with $\bar{g}=0$ is absolutely unstable
with respect to spontaneous formation of a superstructure ($\bar{g}\neq 0$). In the
interval $\alpha_0<\alpha<\alpha_c$, $\alpha_c\approx 1.14\alpha_0$, the state with
$\bar{g}=0$ become metastable; the system still has the ground state with an
equilibrium superstructure. At $\alpha_c<\alpha<\alpha_1$, where $\alpha_1\approx
1.25\alpha_0$, the state with $\bar{g}\neq 0$ becomes metastable. Finally, at
$\alpha>\alpha_1$ Eq.(\ref{eq:f26}) has no solutions with $\bar{g}\neq 0$.

For instance, let us write the expressions for the equilibrium $\bar{g}$ and
$\bar{\omega}_m$ values in the region specified by the conditions $g_{1m}\gg \mu >0$.
Note that the condition $\bar{g}\gg \mu$ is more strong than the condition
$\alpha<\alpha_0$. Inserting $\left|\partial\Omega_g/\partial g\right|$ from
(\ref{eq:f27}) into (\ref{eq:f26}), we obtain
\begin{equation}
\bar{g}\approx \frac{A}{2t\alpha^2};\;\;\;\bar{\omega}_m\sim \frac{1}{\alpha},
\label{eq:f30}
\end{equation}
where $A\sim 1$. The relationship between $\bar{\omega}_m$ and $\alpha$ corresponds to
the obvious fact, namely, increasing the characteristic level spacing favours the
stability of the state with a superstructure.

It is essential that in this state (stable or metastable), the fermion atoms on the
Fermi surface are localized. This implies that that if at a fixed $\alpha$ values the
occupancy of the band is decreased, the transition to the state with a superstructure
occurs jumpwise at
\begin{equation}
\mu=\mu_c=\frac{1}{\alpha^2 t}. \label{eq:f29}
\end{equation}

At $-2g<\mu<0$, there exist the discrete levels only. In this region the state with
$\bar{g}\neq 0$ exists regardless of the $\alpha$ value, and the fermions are
localized on the discrete levels. In this sense the $\mu=0$ value determines the
maximally unstable state of the system. The particle number $N^{(0)}_{c}=0.2 N_c$
corresponds to this state.

{\bf 3.} Let us now discuss instability in a two-dimensional system. For the sake of
simplicity, we assume here that the periodical potential is along one of the principal
axes for the sake of simplicity. Knowing the variation $\delta\rho=\rho-\rho_0$ of the
density of states (\ref{eq:f017})-(\ref{eq:f18}),we can easy find the variations of
thermodynamic quantities. With logarithmic accuracy,
\begin{equation}
\begin{split}
&\delta\Omega_g\approx \delta\mathcal{E}_g,\;\; \delta N(\mu)=-\mu\delta\rho(\mu)\\
&\delta\mathcal{E}_g\approx\rho_{00}g^2\ln\left(\frac{t}{|\mu|+2g}\right)
\end{split}
\label{eq:f31}
\end{equation}
($\mu$ is measured from the middle of the initial band). A decrease in the total atom
energy in the presence of a wave may lead to spontanous formation of a superstructure
as in the one-dimensional case. But because $\delta\mathcal{E}_g$ is a smooth function
of $g$ in the two-dimensional case, the transition occurs in a way similar to a
second-order phase transition. Substituting (\ref{eq:f31}) into the Landau functional
(\ref{eq:f25}), we arrive at the instability condition in the form
\begin{equation}
2\rho_{00}\ln\left(\frac{t}{|\mu|+2g}\right)>\alpha, \label{eq:f32}
\end{equation}
which met only when the $\alpha$ value is low enough. For instance, when the band is
half-filled, at $|\mu|=0$, an equilibrium superstructure is of the scale
\begin{equation}
g_c\approx t \exp(-t\alpha), \label{eq:f33}
\end{equation}

For $|\mu|>g_c$ the homogeneous state is stable. In the state with a superstructure
the particles move freely along one axis, while along the other the motion is
localized.

\section{Concluding remarks}

{\bf 1.} We have found the phase states of one- and two-dimensional Fermi gas in an
optical lattice under interaction with a long-wave perturbation at $T=0$. The wave
vector of the perturbation satisfies the conditions either (\ref{eq:f1}) or
(\ref{eq:f12}). The existence regions of the states are determined by the relations
between the following parameters: the amplitude of the periodical perturbation ($g$),
the initial band wigth $t$, the chemical potential value $\mu$ for fermionic atoms, as
well as the $\alpha_0$ value defining the tangent to the $\partial\Omega_g/\partial g$
curve. The $\alpha_0$ value is determined by Eq.(\ref{eq:f28}). The $\mu$ value is
equivalent to the Fermi energy.

As is shown above, in the one-dimensional system the quantum phase transition to the
groung state with an equilibrium superstructure occurs for low particle numbers when
$\mu<\mu_c$. The $\mu_c$ value is given by the expression (\ref{eq:f29}). In new state
fermions are localized at the Fermi level. In the other words, we may perform the
transition to the state with $\bar{g}\neq 0$ decreasing the particle number at a fixed
$\alpha$ value.

It is worth noting that decreasing the particle number leads to increase of the
$\alpha_0$ value. It makes possible the transition to the state with a superstructure
at $\mu>\mu_c$ provided $\alpha<\alpha_0$.

When $N\leq  N^{(0)}_{c}$, $N^{(0)}_{c}= 0.2 N_c$, the fermionic atoms are localized
for any values of another parameters.

In the two-dimensional system the state with an equilibrium superstructure exists in
the case of a nearly half-filled band. New groung state is realized under condition
$|\mu|<g_c$ (we recall that $\mu$ is measured from the middle of the initial band).
The $g_c$ value is determined in Eq.(\ref{eq:f33}). According to Eqs. (\ref{eq:f33}),
the $g_c$ value depends on $t$ and $\alpha$ exponentially. For this reason, the
condition $|\mu|<g_c$ is rather stringent, and can be satisfied for small enough
values of $t$ and $\alpha$.

In the states with $\bar{g}=0$ the long-wave perturbation generates the narrow peaks
of the density of states near the Fermi level, as is shown in
Eqs.(\ref{eq:f15})-(\ref{eq:f17})  and (\ref{eq:f18}). This may provide the essential
increasing the superfluid temperature in very anisotropic optical lattices.

{\bf 2.} It is well known \cite{VG}, the main anomalous properties of the "old"
superconductors with a quasi-(one,two)-dimensional subsystems of carriers can be
explained only by assuming that near the Fermi surface the density of states exhibits
a narrow peak apparently generated by closeness to the structural transition. Тhis is
caused by the essential dependence of the superconducting transition temperature on
the density of states.

For a dilute fermionic gas $k_F|a_s|\ll 1$, $k_F$ is the Fermi momentum. Thus, the
Cooper pairing occurs in a weak coupling regime, and we obtain the ratio of the
superfluid (BCS) temperature $T_s$ to the Fermi temperature $T_F$ in the form:
$$\frac{T_s}{T_F}\sim \exp\left(-\frac{\pi}{2|a_s|k_F}\right)\sim
\exp\Big(\frac{1}{N(0)|U|}\Big)\ll 1,$$ where $N(0)=m_f k_F/(2\pi^2 \hbar^2)$ is the
density of states at the Fermi surface of a noninteracting gas, $U=4\pi\hbar^2
a_s/m_f$.

Let us consider the superfluidity of fermionic atoms in high anisotropic optical
lattices possessing quasi-(one, two)-dimensional subsystems of fermions. Under a
long-wave perturbation the peaks in the density of states obtained above must give the
essential contribution into the superfluid temperature. As a result, we have both
increasing the $T_s$ value in comparison with the case without the perturbation and
the characteristic dependence of $T_s$ on the Fermi level position (or, what is the
same, on the particle number in lattice).

First, we can start with the situation of nearly ideal quasi-(one, two)-dimensional
gas interacting with a long-wave perturbation of an optical lattice. In this system
the peaks of the density of states near the Fermi level are described by the
expressions (\ref{eq:f17}) and (\ref{eq:f18}). Recall that the peak widths $\sim 2g$
are much smaller than the band width $t$.

Second, we add the degenetrate fermionic gas in an another spin state. In that way,
there may be "switched" the s-scattering with negative scattering length between two
spin components. At $t\gg |U|$ Cooper pairing occurs in a weak coupling regime. In
this regime the densities of states (\ref{eq:f15})-(\ref{eq:f17})  and (\ref{eq:f18})
enter to the $T_s$ temperature. The expressions (\ref{eq:f17}) and (\ref{eq:f18})
imply that the superfluid temperature has the narrow maximum at $\mu\approx 2g$ in the
quasi-one-dimensional case as well as the "smeared" maximum at $-2g<\mu<2g$ in the
quasi-two-dimensional system.

Now let us to discuss briefly the dependence $T_s(\mu)$ when the particle number
changes in the system.

In the quasi-one-dimensional case fermionic atoms are localized at the Fermi level
provided the particle number is relatively small, that is,  $\mu<\mu_c$ at a fixed
$\alpha$. When the particle number increases, and it turns out that $\mu>\mu_c$,
fermions are delocalized. Thus, the dielectric-superconductor transition occurs when
the particle number correspons to the relation $\mu=\mu_c$.

Eq.(\ref{eq:f18}) implies, that in the quasi-two-dimensional case the $T_s(\mu)$ curve
has a break in the slope at the points $\mu=-2g,2g$ near the Fermi energy. It is worth
noting that the state with the fermions localized in the direction of the periodical
potential is realized at $|\mu|<g_c$. For this reason, the $T_s(\mu)$ temperature
decreases within this region.

Thus, both long-wave fluctuations of the order parameter in the neighborhood of an
another quantum phase transition and long-wave perturbations of polarization (as well
as magnitization) generated by external drives can be favourable for the superfluid
state of fermionic atoms.

\section {Acknowledgments.}
This work is supported by the Russian Foundation for Basic Researches.


\end{document}